\documentclass[useAMS]{mn2e}
\usepackage{amsmath}
\usepackage{textcomp}
\usepackage{amssymb}
\usepackage[pdftex]{graphicx}
\usepackage{amssymb}

\usepackage[margin=.8in, paperwidth=8.5in, paperheight=11in]{geometry}
\usepackage{epsfig}
\usepackage{graphicx}
\usepackage{caption}
\begin{document}

\title{Irradiation of Astrophysical Objects - \\ SED and Flux Effects on Thermally Driven Winds}
\author[S. Dyda, R. Dannen, T. Waters, D. Proga]
{\parbox{\textwidth}{Sergei~Dyda\thanks{sdyda@physics.unlv.edu}, Randall Dannen, Tim Waters, Daniel Proga}\\
Department of Physics \& Astronomy, University of Nevada Las Vegas, Las Vegas, NV 89154 
}

\date{\today}
\pagerange{\pageref{firstpage}--\pageref{lastpage}}
\pubyear{2016}

\label{firstpage}

\maketitle

\begin{abstract}
We develop a general method for the self consistent calculation of the hydrodynamics of an astrophysical object irradiated by a radiation field with an arbitrary strength and spectral energy distribution (SED). Using the XSTAR photoionization code, we calculate heating and cooling rates as a function of gas photoionization parameter and temperature for several examples of SEDs: bremsstrahlung, blackbody, hard and soft state XRBs, Type 1 and Type 2 AGN. As an application of our method we study the hydrodynamics of 1-dimensional spherical winds heated by a uniform radiation field using the code 	\textsc{Athena++}. We find that in all cases explored a wind settles into a transonic, steady state. The wind evolves along the radiative heating equilibrium curve until adiabatic cooling effects become important and the flow departs from radiative equilibrium. If the flow is heated very rapidly, for example as in a thermally unstable regime, the corresponding column density of gas is low. Perhaps one of the most intriguing results of our work is the two stage acceleration of the wind that happens when there are two thermally unstable regions and the flux is relatively high. The efficiency with which the radiation field transfers energy to the wind is dependent on the SED of the external source, particularly the relative flux of soft X-rays.  These results suggest that detailed photoionization calculations are essential not only to predict spectra but also to properly capture the flow dynamics.        
 
\end{abstract}

\section{Introduction}

The gas dynamics of many astrophysical systems is affected by radiation from an external source. Planets and moons are irradiated by their host star (e.g., Johnstone, Hollenbach \& Bally 1998; Alexander, Clarke \& Pringle 2006; Owen et al. 2010) and white dwarves can be evaporated by their companion (e.g., Liu, Meyer \& Meyer-Hofmeister 1995). Likewise outflows are radiatively driven by AGN (e.g., Begelman, McKee \& Shields 1983 henceforth BMS83)and X-ray binaries (e.g., Basko \& Sunyaev 1973; Basko et al. 1977; London, McCray \& Auer 1981; Luketic et al. 2010) and the outer layers of accretion discs are irradiated by their inner regions. It is therefore important to develop very general yet robust numerical methods for simulating the heating of hydrodynamic systems by external radiation fields. We developed such a method using the photoionization code XSTAR (Bautista \& Kallman 2001) and MHD code 	\textsc{Athena++} (Gardiner \& Stone 2005, 2008).   

One possible mechanism for launching winds observed in AGNs and YSOs is thermal driving (see Lamers \& Cassinelli 1999 and references therein). In a thermally driven wind, the thermal energy of the gas comes to dominate the gravitational binding energy of the central object and the wind launches. If the thermal energy at the base is insufficient to launch the wind this scenario can still operate provided an external radiation source, such as the central object, heats the gas to sufficiently high temperatures. 

The innermost regions of AGN and XRBs are known to be strong emitters of X-rays. These can heat the gas and provide the thermal energy required for wind launching. An interesting question therefore is how the SED of the external source affects the outflowing wind? This has been previously explored in the context of disc photoevaporation models (e.g., Bally \& Scoville 1982; Hollenbach, Johnstone \& Shu 1993; Owen, Clarke \& Ercolano 2012 and references therein). There are important observational implications because radiative heating can provide a link between the X-ray spectra emitted by the central region and the lines produced further out in the wind.

In the most basic physical picture, gas at the base of the wind is cold and slow and thermal processes dominate over the dynamical processes. The external radiation source heats the wind and the gas evolves along the radiative equilibrium curve. Once the gas is hot enough, dynamical processes such as adiabatic expansion become important and the gas departs from equilibrium. Precisely when this occurs depends on the flux of incident radiation. This is observationally interesting because the line emission and absorption depends on the temperature and ionization state of the wind.

Real systems are complex (e.g., they are non-spherical, could be optically thick and rotate). However, to illustrate our method and to isolate the effects of different SEDs on thermal winds we study a simple problem of 1-dimensional spherical winds in the optically thin limit.  Observed SEDs are typically a superposition of more basic spectra, such as a low temperature blackbody and high frequency powerlaw. Therefore we first consider more basic SEDs such as bremsstrahlung and blackbody to elucidate the basic physics at play. We then use observationally motivated SEDs from Type 1 and Type 2 AGN and soft state and hard state XRBs and generate the corresponding heating rates using the XSTAR photoionization code as a function of gas temperature, T and ionization, $\xi$. These heating rates are used to find spherically symmetric wind solutions using the MHD code 	\textsc{Athena++}. These are used to determine the effects of radiation flux and driving SED on the observationally interesting wind properties such as mass flux and absorption measure distribution.  

The structure of this paper is as follows. In Section \ref{sec:numerical_methods}, we describe our numerical methods including the photoionization code XSTAR, the MHD code 	\textsc{Athena++}  and our implementation of the heating/cooling module. In Section \ref{sec:general_results}, we apply these methods to find spherically symmetric winds thermally driven by different types of external radiation fields and explore their observational features. In Section \ref{sec:discussion}, we summarize our findings and discuss the limitations and future applications of these methods, in particular their relevance to 2D and 3D problems.   

\section{Numerical Methods}
\label{sec:numerical_methods}
We are interested in performing hydrodynamic simulations where the radiative heating/cooling is calculated self-consistently using a photoionization code. We implement a heating/cooling module into 	\textsc{Athena++} where the energy equation is sourced by a net cooling rate $\mathcal{L}$ that accounts for the radiative heating and cooling of the gas by an external source. The rate is calculated self-consistently using the \textsc{XSTAR} photoionization code as a function of the temperature $T$ and ionization parameter $\xi$ calculated by 	\textsc{Athena++}. Photoionization calculations are computationaly expensive so we precalculate $\mathcal{L}(\xi,T)$ on a grid $(\xi_i, T_j)$. We then interpolate between points on this grid within 	\textsc{Athena++} to calculate the net heating. Below we briefly describe our \textsc{XSTAR} simulation, 	\textsc{Athena++} simulation and implementation of the heating/cooling term. 

\subsection{Photoionization - \textsc{XSTAR}}

To accurately model photoionization processes we used version 2.3 of the code \textsc{XSTAR} (Bautista \& Kallman 2001). By assuming a certain SED of incident flux of radiation on a box of gas containing known species of gas, \textsc{XSTAR} calculates the heating/cooling rate as a function of temperature and ionization fraction in the gas. For a review of methods for modelling photoionized plasmas see Kallman (2010).

For ease in comparing results for different SEDs $\mathcal{F}(E)$ we relate the mean photon energy $\langle h \nu \rangle$ to the X-ray temperature $T_X$ via   
\begin{equation}
\langle h \nu \rangle = k_b T_X,
\end{equation}
where $k_b$ is the Boltzman constant and we use $h \nu_0 = 0.1 \ \rm{eV}$ as the low energy X-ray cutoff. This ensures that at large ionization fractions, when Compton processes dominate, the equilibrium curves will match up for all SEDs. Though we find that the gas never reaches this Compton temperature of $T_{\rm{IC}} = 1/4 T_X$ in transonic solutions, our goal was to have the greatest possible uniformity across our runs.

Within our \textsc{XSTAR} simulation we assume the heating/cooling in the gas is due to Compton, X-ray, Bremsstrahlung and line processes. Our main model dependent assumptions are \textbf{1)} Composition of the gas. \textbf{2)} Available atomic transitions. \textbf{3)} Source SED. 

We use the elemental abundances described in Verner, Barthel \& Tytler (1994) where elements with abundances $< 10^{-6}$ of hydrogen were set to zero. We found that gas composition can play an important role, namely in determining the equilibrium curve. The AGN cases exhibited an unphysical downwards dip at low ionization parameter. For these cases we therefore used the abundances in Lodders, Palme \& Gail (2009). We expect most variations due to composition to occur at low ionization/temperature where line processes are important. Because we are mainly interested in the relationship between the equilibrium curve and the wind we chose to leave a detailed study of the gas composition on the equilibrium curve for future work.

Comprehensive atomic data is crucial for determining the thermal properties of gases hotter than $~10^{4} \rm{K}$. \textsc{XSTAR}v2.35 incorporates over 200 000 lines for gas at a given $(\xi,T)$. As photoionization codes improve by incorporating better atomic data and more comprehensive lists on atomic transitions our methods for calculating net heating will only improve.

Our main focus within this study is how the source SED affects the net heating rate and in turn affects the driven wind. We describe our philosophy regarding choice of SEDs in \ref{subsub:SED}.

\subsection{Hydrodynamics - 	\textsc{Athena++}}

\subsubsection{Basic Equations}
The basic equations for single fluid hydrodynamics are
\begin{subequations}
\begin{equation}
\frac{\partial \rho}{\partial t} + \nabla \cdot \left( \rho \mathbf{v} \right) = 0,
\end{equation}
\begin{equation}
\frac{\partial (\rho \mathbf{v})}{\partial t} + \nabla \cdot \left(\rho \mathbf{vv} + \mathbf{P} \right) = - \rho \nabla \Phi,
\end{equation}
\begin{equation}
\frac{\partial E}{\partial t} + \nabla \cdot \left( (E + P)\mathbf{v} \right) = -\rho \mathbf{v} \cdot \nabla \Phi - \rho \mathcal{L}(\xi,T),
\label{eq:energy}
\end{equation}
\label{eq:hydro}%
\end{subequations}
where $\rho$ is the fluid density, $\mathbf{v}$ the velocity, $\mathbf{P}$ a diagonal tensor with components P the gas pressure, $\Phi = -GM/r$ is the gravitational potential of the central object and $E = 1/2 \rho |\mathbf{v}|^2 + \mathcal{E}$ is the energy where $\mathcal{E} =  P/(\gamma -1)$ is the internal energy. $\mathcal{L}$ is the net cooling rate described in Section \ref{sub:heating} and is assumed to be a function of temperature $T$ and ionization parameter
\begin{equation}
\xi = \frac{4 \pi \mu m_p F_X}{\rho},
\label{eq:xi_def}
\end{equation}
where $F_X$ is the X-ray flux which we assume to be uniform, $\mu$ is the mean molecular weight and $m_{\rm{p}}$ is the proton mass.  We take an equation of state $P = \rho^{\gamma}$ where $\gamma = 5/3$. The isothermal sound speed is $a^2 = P/\rho$ and the adiabatic sound speed $c_s^2 = \gamma a^2$.  We can compute the temperature from the internal energy via $T = (\gamma -1)\mathcal{E}\mu m_{\rm{p}}/\rho k_{\rm{b}}$.

\subsubsection{Simulation Parameters}
\label{subsub:simulation_parameters}

We choose parameters for our different runs to be able to consistently compare simulations between different classes of SEDs. A useful measure for thermally driven winds is the hydrodynamic escape parameter $\rm{HEP} = -\Phi/c_s^2 = GM\mu m_p/r\gamma k_b T$, which measures the ratio of gravitational to thermal energy. At the base of the wind, for $HEP \gg 1$ the gas is gravitationally bound and no thermally driven outflow is expected. However for $HEP \leq 10$ thermally driven hydrodynamic winds will be produced (Stone \& Proga 2009). We \textbf{fix the} central object mass $M = M_{\odot}$ to help in comparing results across our different runs. Our results could be applied to systems with different masses, for instance AGN systems, if the relevant length (see eq \ref{eq:compton_radius}) and ionizing flux (see eq \ref{eq:F_critical}) are scaled appropriately. At the base of the wind, for most runs, we set the initial temperature $T_* = 2.5\times 10^{4} \rm{K}$ and set $\rm{HEP} = 5$. This sets the inner radius $r_* = 4.605 \times 10^{12} \rm{cm}$. The exception is for the Blackbody SED, because the equilibrium curve is thermally unstable at this temperature, we set $T_* = 1.7 \times 10^{4} \rm{K}$ which corresponds to a $HEP = 7.4$ and is still in the regime where winds can be thermally driven. These temperatures ensure that we are on the flat part of the S curve in the cold phase for all SEDs and also slightly away from any thermally unstable region. By initializing the gas in equilibrium near the star, the ionisation parameter $\xi_*$ is fixed for each SED. 

We explored the effects of varying $\xi_*$ and $T_*$ values along the base of the radiative equilibrium S curve. For the mBl case with $F = 10^{-3} F_{\rm{cr}}$  we compared our fiducial $T_* = 25 000 \rm{K}$ case with $T_* = 19 000 \rm{K}$ and $T_* = 52 000 \rm{K}$ cases. We found that at large radii the velocity differed by $0.1 \%$ and $1\%$ and temperature varied by $1 \%$ and $6 \%$ respectively. We concluded that the outflows at large radii are largely insensitive to position on the radiative equilibrium curve, as one would expect for transonic flows.

A physically relevant length scale for this problem is the Compton radius
\begin{equation}
R_{\rm{IC}} = \frac{GM\mu m_p}{k T_{IC}} = 9.648 \times 10^{9} \left( \frac{M}{M_{\odot}}\right) T_{\rm{IC,8}}^{-1} \ \rm{cm},
\label{eq:compton_radius}
\end{equation}
which is the radius at which gas at $T_{IC}$ has internal energy equal to its gravitational potential energy. Above we have set $T_{\rm{IC,8}} = T_{\rm{IC}}/10^8 \rm{K} $. For a stellar mass central object and our fiducial choice of Compton temperature this corresponds to a Compton radius $R_{IC} = 3.5 \times 10^{10} \rm{cm}$.
  
Near the Compton temperature, radiative processes will be dominated by Compton heating and cooling. The Compton heating rate can be used to define a critical radiation flux 
\begin{equation}
\begin{aligned}
F_{\rm{cr}} =& \frac{m_e c^2}{4 \sigma_T} \frac{HEP}{GM} \left(\frac{\gamma k_b}{\mu m_p}\right)^{3/2} T_* \ T_{\rm{IC}}^{1/2}  \\ =& 4.04 \times 10^{12} \left( \frac{HEP}{5}\right) \left( \frac{M}{M_{\odot}}\right) T_{\rm{*,4}} \  T_{\rm{IC,8}}^{1/2} \ \rm{erg \ cm^{-2} \ s^{-1}},
\end{aligned}
\label{eq:F_critical}
\end{equation}
where $T_{\rm{*,4}} = T_{\rm{*}}/ 10^4 \ \rm{K}$. This corresponds to the flux at which X-rays can heat the gas to $T_{\rm{IC}}$ via Compton processes in the sound crossing time of $r_*/c_s$. This is the analogue of the critical luminosity of BMS83 (eq 2.12) for a uniform radiation field. We explore simulations where the flux ranges from $F/F_{\rm{cr}} = 10^{-6}-10^{0}$. 

We are interested in wind solutions that have been launched due to heating. Therefore we set the radius $r \gg R_{IC}$ and the gas is initially at $T \ll T_{IC}$ which are favourable conditions for radiative wind launching. By varying the flux we can study the effects of the particular SED on the launched winds. 

Assuming a uniform radiation field allows us to define a fiducial density $\rho_0$ at the base of the wind for each SED via
\begin{equation}  
\rho_* = \frac{4 \pi \mu m_p F_{\rm{cr}}}{\xi_*} \left(\frac{F}{F_{\rm{cr}}}\right) = \rho_0 \left(\frac{F}{F_{\rm{cr}}}\right).
\end{equation}
Because the ionization parameter is fixed, a change in flux is equivalent to a change in density of the gas. A summary of our simulation parameters for all runs are shown in Table \ref{table:run_summary}. In Section \ref{sec:discussion}, we discuss the effects of non-uniform radiation fields, caused by geometric or optical depth effects.

\begin{table*}
\begin{tabular}{l | c | c | c | c | c | c | c | c |}
\hline \hline
SED (Model) & Instability & $T_{\rm{IC}}$ & $R_{\rm{IC}}$ & $F_{\rm{cr}}$ & $\xi_*$ & $T_*$ & $\rm{HEP}$ & $\rho_{0}$  \\ 
&  & $[10^8 \ \rm{K}]$ & $[10^{10} \ \rm{cm}]$ & $[\rm{erg \ cm^{-2} \ s^{-1}}]$ & $\rm{[erg \ cm \ s^{-1}}]$ & $[10^4 \ \rm{K}]$ & & $[10^{-12}\rm{g \ cm^{-3}}]$ \\ \hline \hline
Modified Blondin (mBl)                  & None 		& 0.28 & 3.43 & $5.34 \times 10^{12}$ & 5.24  & 2.50 & 5 & 12.9  \\ 
Blondin (Bl)                            & Isobaric 	& 0.28 & 3.43 & $5.34 \times 10^{12}$ & 4.75  & 2.50 & 5 & 14.2 \\ 
Bremsstrahlung (Brem)	                & Isobaric 	& 0.28 & 3.43 & $5.34 \times 10^{12}$ & 28.36 & 2.50 & 5 & 2.39 \\ \hline
Blackbody (BB)	                        & Isochoric 	& 0.28 & 3.43 & $3.63 \times 10^{12}$ & 1320  & 1.70 & 7.4 & 0.035 \\ \hline \hline
Soft State XRB (XRB1)	                & Isobaric	& 0.28 & 3.43 & $5.34 \times 10^{12}$ & 24.09 & 2.50 & 5 & 2.81 \\
Hard State XRB (XRB2)	                & Isobaric	& 0.75 & 1.29 & $8.74 \times 10^{12}$ & 12.29 &	2.50 & 5 & 9.02 \\ \hline
Type 1 AGN (AGN1)    			& Isobaric	& 1.00 & 0.96 & $1.01 \times 10^{13}$ & 2.16  & 2.50 & 5 & 59.2 \\
Type 2 AGN (AGN2)			& Isobaric	& 1.52 & 0.63 & $1.24 \times 10^{13}$ & 6.47  & 2.50 & 5 & 24.4 \\ 
\hline \hline
\end{tabular}  
\caption{Summary of parameters for different radiation models. There are four families of models - 1) Bremsstrahlung like including an \textsc{XSTAR} model (Brem) and analytic fits based on Blondin (1994) (mBl \& Bl) 2) Blackbody (BB) 3) X-ray binary in soft state (XRB1) and hard state (XRB2) 4) Type 1 (AGN1) \& Type 2 (AGN2) AGN}
\label{table:run_summary}
\end{table*}

\subsubsection{Grid \& Boundary Conditions}
The simulation region extends from $r_* < r < 10 r_*$. By definition of the critical flux, we expect the wind to heat to roughly $T_{\rm{IC}}$ over these length scales. We use a logarithmically spaced grid of $N_r = 120$ points and a scale factor $a_r = 1.09$ that defines the grid spacing recursively via $dr_{n+1} = a_r dr_n$.

At the inner boundary we impose outflow boundary conditions on $\mathbf{v}$ and $E$ while keeping the density fixed at $\rho = \rho_*$ in the first active zone. As a result the ionization parameter $\xi_*$ is fixed in the first active zone. Hence we do not enforce radiative equilibrium in the first active zone except at the initial time because the temperature can evolve away from its initial equilibrium value.  At the outer boundary we impose outflow boundary conditions on $\rho$, $\mathbf{v}$ and $E$.

\subsection{Heating/Cooling}
\label{sub:heating}

%%%%%%%%%%%%%%%%%%%%%%%%%%%%%%%%%%% Observational Spectra
\begin{figure*}
                \centering
                \includegraphics[width=\textwidth]{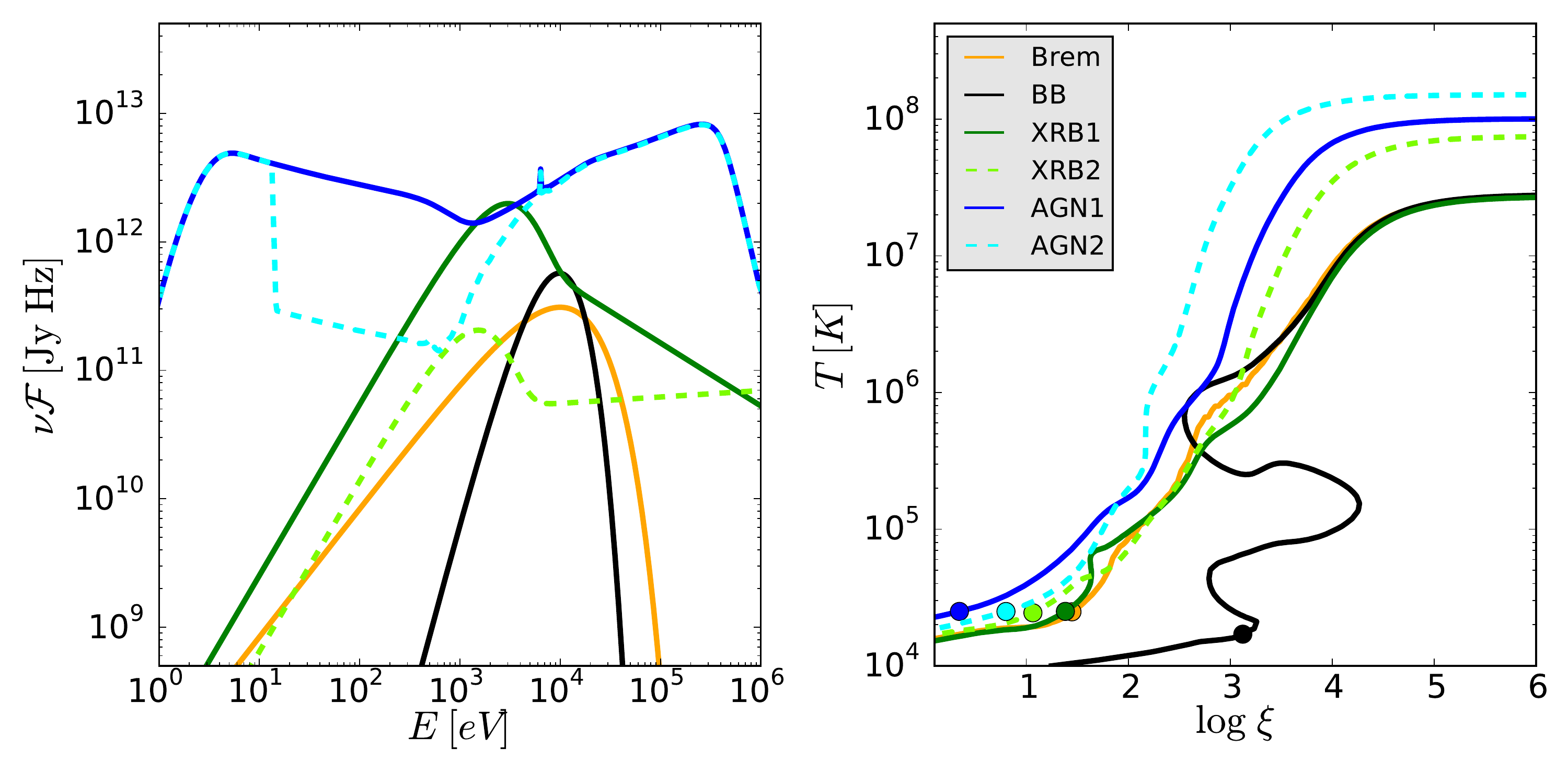}
        \caption{\textit{Left -} SED for modeling ideal Bremsstrahlung (Brem) and Blackbody (BB) radiation and observational SEDs for Type 1 (AGN1) and Type 2 AGN (AGN2) (Mehdipour et al. 2015) and  X-ray binaries in soft (XRB1) and hard (XRB2) states (Trigo et al. 2013). \textit{Right -} Corresponding equilibrium curves generated by \textsc{XSTAR}. The dot indicates the initial conditions at the base of the wind at $T = 1.7 \times 10^{4} ~ \rm{K}$ (BB) and $T = 2.5 \times 10^{4} ~ \rm{K}$ (all other cases) .          
}
\label{fig:observed_SED}
\end{figure*} 
%%%%%%%%%%%%%%%%%%%%%%%%%%%%%%%%%%%

Within 	\textsc{Athena++} we assume that all the microphysics incorporated into \textsc{XSTAR} is captured by a net cooling rate $\mathcal{L}(\xi,T)$ that depends only on the macroscopic gas properties, temperature $T$ and ionization parameter $\xi$. We have implicitly assumed that the microphysical processes occur on much shorter time scales than the dynamical time governing the hydro simulation which justifies this approach. We give a brief description of our 	\textsc{Athena++} heating module in Section \ref{subsub:Athena++} and provide a detailed description along with some numerical testing in Appendix \ref{sec:heating_appendix}. We describe our different choices of SED providing this heating in Section \ref{subsub:SED}.  

\subsubsection{	\textsc{Athena++} Implementation}
\label{subsub:Athena++}  

First we calculate $\mathcal{L}(\xi,T)$ on a logarithmically spaced grid $(\xi_i,T_j)$ of size $200 \times 75$ over the range $5 \times 10^3 \ K < T < 10^8 \ K$ and $10^0 < \xi < 10^8$ using \textsc{XSTAR}. This takes roughly one week running on 12 cores for each SED.   

At every timestep in the hydro simulation we must calculate the rate of change in energy of the gas from radiative processes, the $\rho \mathcal{L}$ term appearing in the energy equation (\ref{eq:energy}). This is related to the heating rate $\Gamma$ and cooling rate $\Lambda$ calculated by \textsc{XSTAR} via
\begin{equation}
\rho \mathcal{L} = \left(\frac{\rho}{\mu m_p}\right)^2 (\Lambda - \Gamma),
\end{equation}
where $[\Gamma] = [\Lambda] = \rm{erg \ cm^{3} \ s^{-1} }$ At every cell center location we calculate the ionization parameter $\xi$ using (\ref{eq:xi_def}). We then bilinearly interpolate over the nearest grid points in $(\xi_i,T_j)$ from our \textsc{XSTAR} grid. A backwards Euler method was used to solve for the heating rate under the assumption that the change in internal energy is due solely to the net heating and not the dynamics of the gas. Using the backwards Euler method ensures that the gas reaches equilibrium smoothly. This is important in this problem because the base of the wind is in equilibrium and this prevents any transients from entering the solution. A detailed description of this method is provided in Appendix \ref{sec:heating_appendix}.

\subsubsection{Driving Radiation Field}
\label{subsub:SED}
Our goal is to use observed SEDs to generate net heating rates to generate realistic hydrodynamics models. We assume this radiation field is uniform throughout the simulation region, as might happen for instance if the central object is surrounded by an optically thin gas and illuminated by a large number of uniform point sources. We first use simpler heating and cooling models, derived from idealized SED's that have equilibrium curves with progresively more complicated features.

We first consider the analytic heating model proposed by Blondin (1994) as a fit to thermal Bremsstrahlung where
\begin{equation}
\begin{aligned}
\Lambda - \Gamma  =& \Bigg[ A_b \times 3.3 \times 10^{-27} \sqrt{T} \\ &+ \left( A_l \times 1.7 \times 10^{-18} \frac{e^{-T_L/T}}{\xi \sqrt{T}} + 10^{-24}\right) \Bigg] \\ &-\Bigg[ A_C \times 8.9\times 10^{-36} \xi (T_x - 4 T) \\ &+ A_X \times 1.5 \times 10^{-21} \frac{\xi^{1/4}}{T^{1/2}}\left( 1 - T/T_X \right) \Bigg] \\
\end{aligned}
\label{eq:heating_cooling}
\end{equation}
where we have used an X-ray temperature $T_X = 1.12 \times 10^{8} \rm{K}$ and a line temperature $T_L = 1.3 \times 10^{5} \rm{K}$. We have chosen the X-ray temperature so that $T_{\rm{IC}} = 2.8 \times 10^{7} K$. Above the terms represent the parametrized heating and cooling from Compton processes ($A_C$), X-rays ($A_X$), Bremsstrahlung ($A_b$) and lines ($A_l$). They depend only on the temperature of the gas T and the ionization parameter $\xi$. 

In our \emph{Modified Blondin} (mBl) model we set $A_C = A_X = A_l = 1.0$, $A_b = 3.9$. This corresponds to the parametrization of ``Model C" in Higginbottom \& Proga (2015) but with an increased X-ray temperature. These parameters were chosen so that the equilibrium curve is free of the thermal instability (TI) (Field 1965). In the \emph{Blondin} (Bl) model we set $A_C = A_X = A_l = A_b = 1.0$ which is the original fit proposed by Blondin (1994) but with increased X-ray temperature. The equilibrium curve is unstable to isobaric perturbations. These analytic models were also used to test our interpolative heating scheme (see Appendix \ref{sec:heating_appendix}). The \emph{Bremsstrahlung} (Brem) model uses an SED of the same name as input to \textsc{XSTAR} for generating net heating rates. The \emph{Blackbody} (BB) model provides a case where the equilibrium curve is thermally unstable to isochoric perturbations (Kallman \& McCray 1982; Buff \& McCray 1974). It can be viewed as either a true blackbody source from optically thin emission or as a proxy for an SED with a low energy cutoff say from absorption by cold gas near the source.

The remaining models are based on SEDs obtained from observations. We consider an \emph{X-ray binary} in a soft state (XRB1) and a hard state (XRB2) (Trigo et al. 2013) and \emph{Active Galactic Nuclei} SEDs from Type 1 (AGN1) and Type 2 (AGN2) AGN (Mehdipour et al. 2015). We note that the shape of the equilibrium curve is highly dependent on the atomic abundances used, in particular at low values of the ionization parameter where the cooling is largely dominated by lines. As an illustration of our method we used AGN spectra for NGC 5548 at two different epochs. However, for other applications like modeling AGN feedback one might want to use more typical AGN spectra (e.g. Sazonov, Ostriker \& Sunyaev 2004; Sazonov et al. 2005).  We show plots of these SEDs and the corresponding equilibrium curves output by \textsc{XSTAR} in Fig. \ref{fig:observed_SED}. We indicate the initial conditions at the base of the wind for each SED with a dot.

\section{Results}
\label{sec:general_results}

At the base of the wind the velocity is subsonic and gravitational energy dominates over thermal energy ($HEP \gtrsim 5$) so basic energy considerations require that energy be added for a transonic wind to launch. This energy is provided by an external radiation field which differs according to each SED.

Irrespective of SED, the flows follow qualitatively similar behaviour - near the base of the wind they are in radiative equilibrium and follow the equilibrium curve until dynamical processes become important and they begin to radiatively heat. We briefly describe the dynamics of the flow in Section \ref{sec:dynamical_variables} for completenesss. However, we are primarily interested in the bulk properties of the flow - the mass flux and efficiency of momentum and energy tranfer to the wind. In particular, we compare how the different radiation fields affect these bulk observable properties in Section \ref{sec:bulk}. We then discuss the spectroscopic properties of these flows by exploring their signatures on the absorption measure distribution Section \ref{sec:AMD}

\subsection{Dynamical Variables}
\label{sec:dynamical_variables}

%%%%%%%%%%%%%%%%%%%%%%%%%%%%%%%%%%% xiT-1_T
\begin{figure*}
                \centering
                \includegraphics[width=\textwidth]{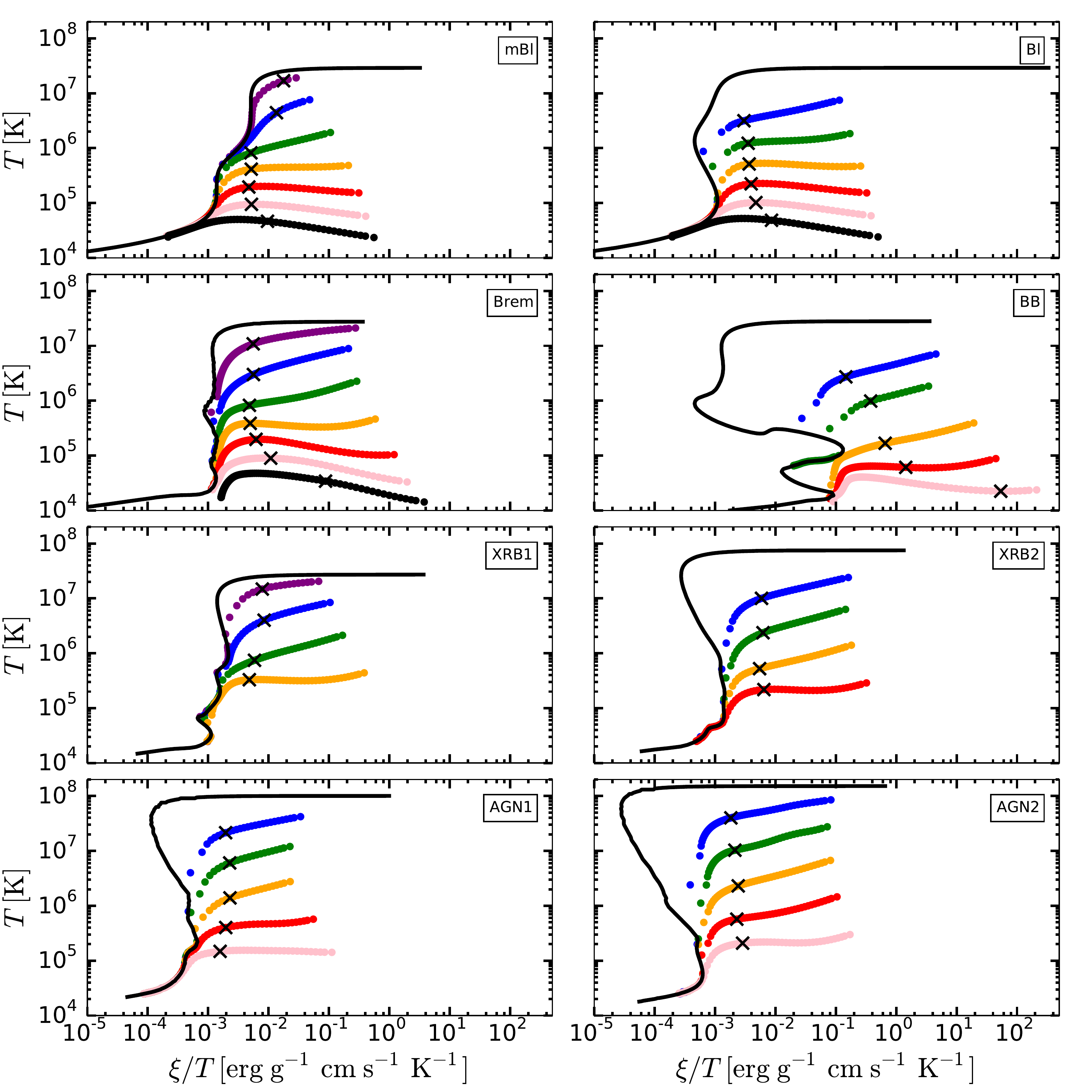}
        \caption{Summary of $\xi/T$-$T$ phase space explored by each radiation field model. For each SED, listed in the top right of each panel (see also Table 1), the color represents a fixed multiple of the critical radiation flux $F/F_{\rm{cr}}$ = $1$ (purple), $10^{-1}$ (blue), $10^{-2}$ (green), $10^{-3}$ (orange), $10^{-4}$ (red), $10^{-5}$ (pink) and $10^{-6}$ (black). The location of the sonic point is indicated by an X           
}
\label{fig:xiT-1_T_summary}
\end{figure*} 
%%%%%%%%%%%%%%%%%%%%%%%%%%%%%%%%%%%

For each SED we find stationary wind solutions for a range of subcritical radiation fluxes. The density is monotonically decreasing and velocity almost always monotonically increasing yielding a constant mass flux at all radii as expected for a stationary flow. The exception is at thermally unstable points where there is a drop in outwards pressure, leading to a dip in the velocity profile and a dip in the mass flux. The flows are qualitativly similar whereby they follow the equilibrium curve near the base of the wind before dynamical processes become important and they begin to heat. We summarize this for all SED models in Fig. \ref{fig:xiT-1_T_summary} where we show the phase space evolution of each solution. Each color corresponds to a fixed fraction of the critical radiation flux for each model $F/F_{\rm{cr}}$ = $1$ (purple), $10^{-1}$ (blue), $10^{-2}$ (green), $10^{-3}$ (orange), $10^{-4}$ (red), $10^{-5}$ (pink) and $10^{-6}$ (black). For larger incident flux, equilibrium is maintained for larger ionizations allowing the flow to reach higher temperatures. Our definition of critical flux (\ref{eq:F_critical}) appears consistent with these results since simulations with critical flux (purple) asymptotically approaches the Compton temperature.  

\begin{figure}
                \centering
                \includegraphics[width=0.45\textwidth]{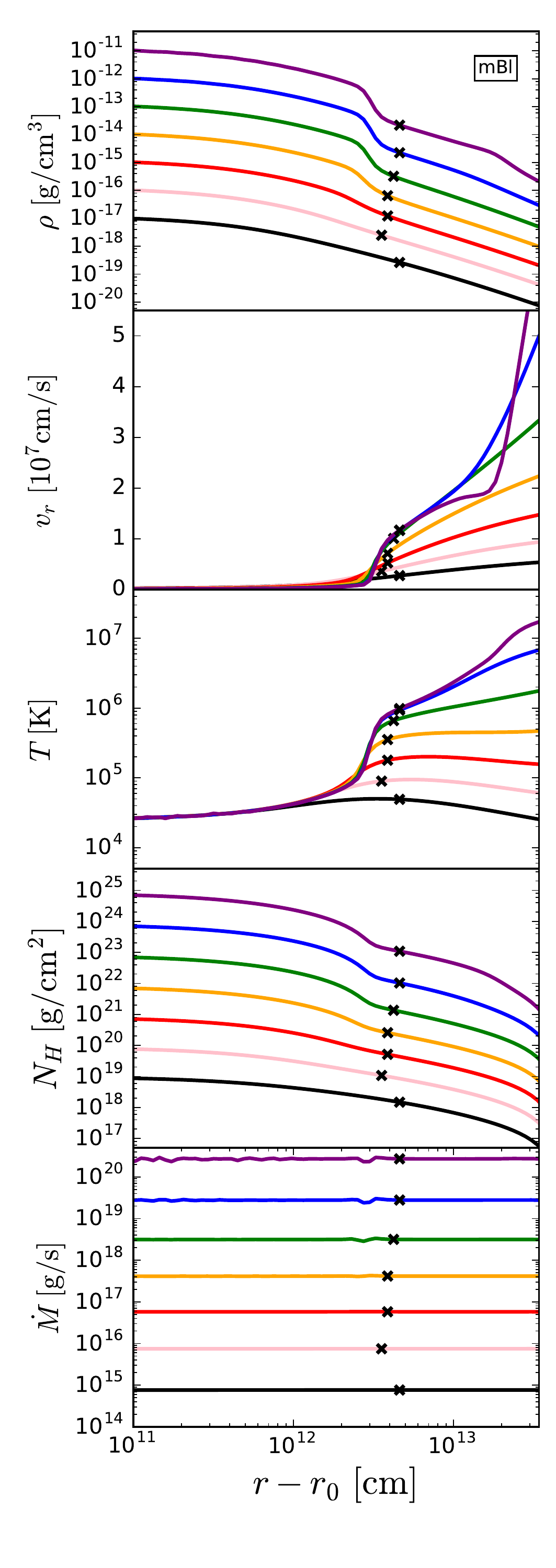}
        \caption{Summary of dynamical variables, density $\rho$, velocity $v_r$, temperature $T$, column density $N_H$ and mass flux $\dot{M}$ for the mBl case. We indicate the location of the sonic point with an X. We omit $10^{8} < r - r_0 < 10^{11}$ where the dynamical variables are nearly constant to focus on the regions near the sonic point.          
}
\label{fig:dynamical_summary}
\end{figure}

As the flow accelerates adiabatic cooling becomes an important contributor to the energy balance, namely the gas temperature is not determined by the balance between radiative heating and cooling only but rather radiative heating, radiative cooling and adiabatic cooling. Consequently, the gas temperature is below the radiatve equilibrium temperature. When flux is sufficiently high (e.g. see the purple curves corresponding to the critical flux) the wind can undergo a second phase of such acceleration (Fig. \ref{fig:dynamical_summary}). When this occurs there is a large spike in the heating rate $dE/dr$ with most of the energy going into thermal energy. This thermal energy is used by the flow to first exit the gravitational potential well. The kinetic energy then dominates over the gravitational energy and thermal energy goes to further increase the velocity of the wind. At large distances the flow is well developed with total energy equipartitioned between thermal and kinetic energy. We note in particular that at the sonic point the rate of heating $q = dQ/dr = \mathcal{L}/v_r $ satisfies $d \ln q/ d \ln r < 1/2$ (see Fig. \ref{fig:energy_summary}), as expected from general energetic arguments for externally heated transonic winds (Holzer \& Axford 1970; Lamers \& Cassinelli 1999). 

%%%%%%%%%%%%%%%%%%%%%%%%%%%%%%%%%%% dynamical summary
%%%%%%%%%%%%%%%%%%%%%%%%%%%%%%%%%%% 

\subsection{Energetics}

%%%%%%%%%%%%%%%%%%%%%%%%%%%%%%%%%%% Observational Spectra
\begin{figure*}
                \centering
                \includegraphics[width=0.95\textwidth]{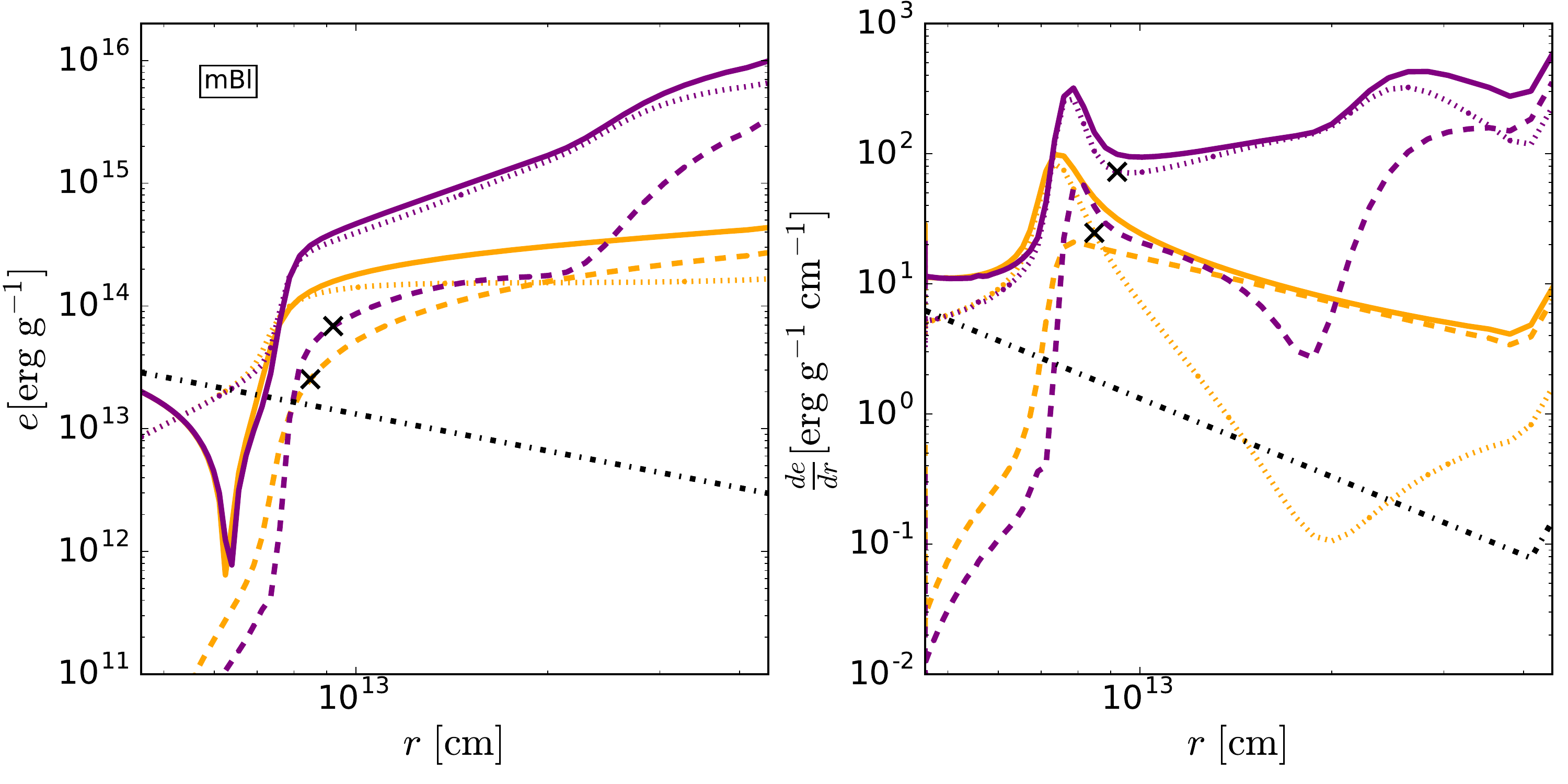}
        \caption{\textit{Left -} Kinetic $e_{\rm{k}}$ (dashed), thermal $e_{\rm{th}}$ (dotted), gravitational $|e_{\rm{grav}}|$ (black dash-dot) and total energy $|e_{\rm{tot}}|$ (solid) as a function of position in the wind for the mBl case with flux $F = F_{\rm{cr}}$ (purple) and $F = 10^{-3 }F_{\rm{cr}}$ (orange)  \textit{Right -} Corresponding $de/dr$ in the wind. Unlike the dynamical variables in Fig. \ref{fig:dynamical_summary}, we plot these as a function of r. At the base of the wind the thermal energy follows the equilibrium curve and most of the change in energy of the wind is from the thermal component. This peaks when the kinetic energy and magnitude of gravitational energy are equal and the flow becomes supersonic. In the subcritical case, the wind becomes nearly isothermal and additional energy is converted to kinetic energy. In the critical case, the wind experiences a second phase of acceleration beyond the sonic point, before also becoming kinetic energy dominated.            
}
\label{fig:energy_summary}
\end{figure*} 
%%%%%%%%%%%%%%%%%%%%%%%%%%%%%%%%%%% 

The physics of radiatively heated winds can be understood by considering the energetics of the wind. Generally, there is a competiton between the thermal energy $e_{\rm{th}}$ and gravitational energy $e_{\rm{grav}}$ per unit mass. If gravitational energy dominates we have an accreting Bondi solution, but if thermal energy dominates then we have a thermal wind solution (Parker 1958). Because we are interested in studying the effects of heating, we considered initial conditions where the ratio of gravitational to thermal energy, as measured by $(\gamma - 1) HEP \gtrsim 1$. Therefore, in order to avoid an accreting Bondi solution, $e_{\rm{th}}$ must come to dominate the energy of the gas.

Suppose we have reached a stationary solution $\partial/\partial t = 0$, then we may express (\ref{eq:hydro}) as the Bernoulli function (e.g Lamers \& Cassinelli 1999)

\begin{equation}
\frac{d}{dr} \left(\underbrace{\frac{v_r^2}{2}}_{e_{\rm{kin}}} + \underbrace{\frac{\gamma}{\gamma -1} \frac{k_b T}{\mu m_p}}_{e_{\rm{th}}} - \underbrace{\frac{GM}{r}}_{e_{\rm{grav}}} \right) = \underbrace{-\frac{\mathcal{L}}{v_r}}_{\frac{dQ}{dr}}.
\label{eq:dedr}
\end{equation} 
We may interpret each term on the left hand side as telling us the change in energy per unit mass of the wind in the kinetic, thermal and gravitational energy as we move outwards in the wind. The radiative heating $\mathcal{L}$ acts as a source of heat $dQ$. We ensure that the energy budget is satisfied by plotting both sides of (\ref{eq:dedr}) which is possible because we have an analytic expression for the heating.

In the lefthand panel of Fig. \ref{fig:energy_summary} we plot $e_{\rm{kin}}$, $e_{\rm{th}}$, $|e_{\rm{grav}}|$ and the total energy $|e_{\rm{tot}}|$ for the representative cases $F = 10^{-3}F_{\rm{cr}}$ and $F = F_{\rm{cr}}$ for the mBl SED. In the right hand panel of Fig. \ref{fig:energy_summary}, we plot $de_{\rm{kin}}/dr$, $de_{\rm{th}}/dr$, $|de_{\rm{grav}}/dr|$ and $|de_{\rm{tot}}/dr|$. Since we are examining steady state solutions this allows us to see where energy is injected into the flow.

At small radii the thermal energy follows the equilibrium curve. When the thermal energy approximately equals the magnitude of gravitational energy the total energy is approximately zero (indicated by the spike in $|e_{\rm{tot}}|$). Between this point and the sonic point energy is primarilly injected into the thermal component. This is where the thermal component falls off the radiative equilibrium curve, since the flow velocity is non-negligible and adiabatic cooling dominates over radiative cooling.

\subsection{Bulk Properties}
\label{sec:bulk}

%%%%%%%%%%%%%%%%%%%%%%%%%%%%%%%%%%% Efficiencies
\begin{figure*}
                \centering
                \includegraphics[width=\textwidth]{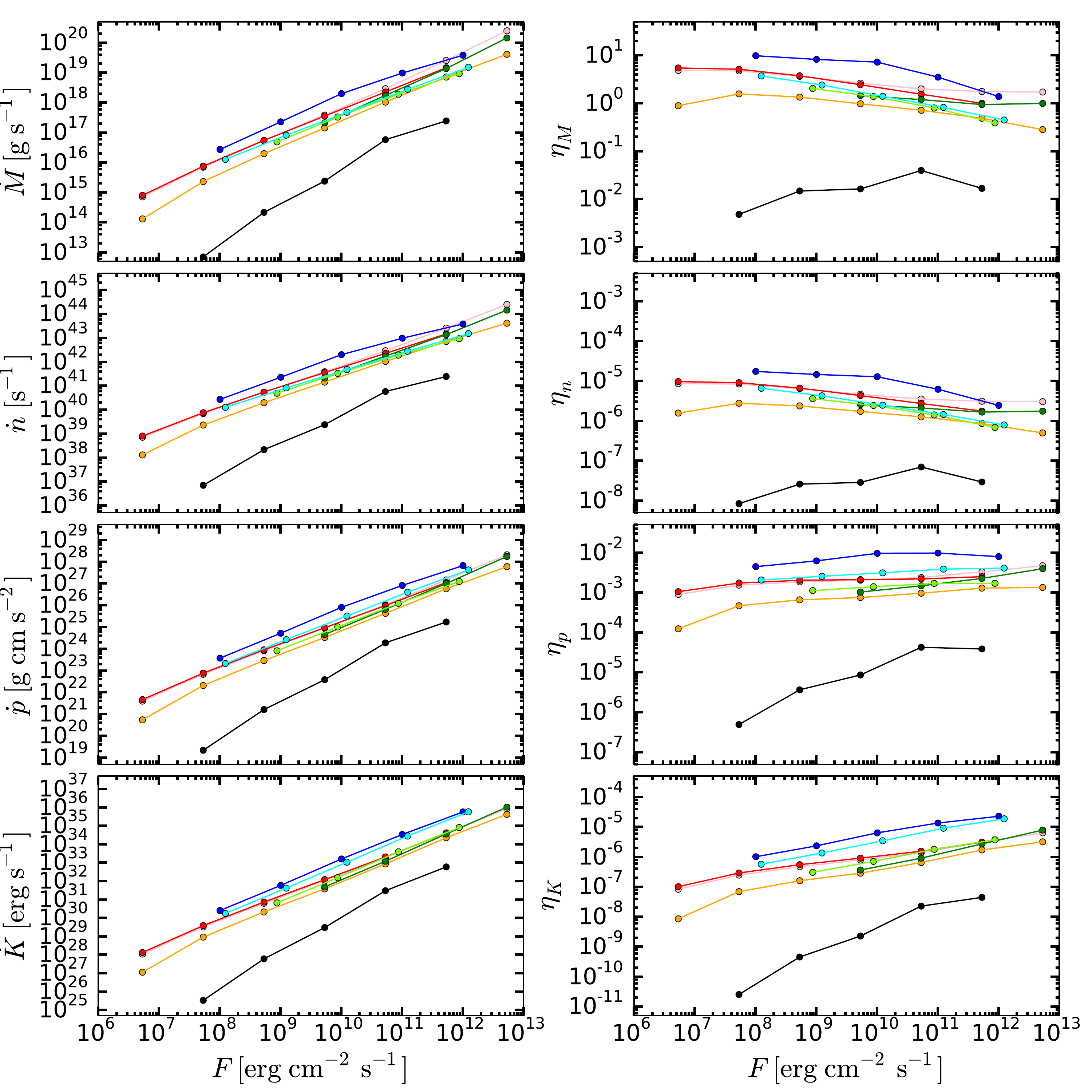}
        \caption{Mass flux $\dot{M}$, particle flux $\dot{n}$, momentum flux $\dot{p}$ and kinetic energy flux $\dot{K}$ at the outer boundary and the corresponding efficiencies $\eta$ as a function of radiative flux F for different SEDs - mBl (pink), Bl (red), Brem (orange), BB (black), XRB1 (green), XRB2 (light green), AGN1 (blue), AGN2 (cyan). BB (black) is an outlying model because the SED has a deficit of $\sim 10^{5}$ in soft photons relative to the other models.           
}
\label{fig:efficiency_summary}
\end{figure*} 
%%%%%%%%%%%%%%%%%%%%%%%%%%%%%%%%%%%

We are interested in how the bulk properties of the flow are affected by the radiation field. In particular, we are interested in the flux of mass, particles, momentum and kinetic energy at the outer boundary and the corresponding efficiencies
\begin{subequations}
\begin{equation}
\dot{M} = 4 \pi \rho v_r r^2 \hspace{1cm} \eta_{M} = \frac{\rho v_r c^2}{F},
\end{equation}
\begin{equation}
\dot{n} = 4 \pi \frac{\rho v_r r^2}{\mu m_p} \hspace{1cm} \eta_{n} = \frac{\rho v_r \langle hf \rangle}{F \mu m_p},
\end{equation}
\begin{equation}
\dot{p} = 4 \pi \rho v_r^2 r^2 \hspace{1cm} \eta_{p} = \frac{\rho v_r^2 c}{F},
\end{equation}
\begin{equation}
\dot{K} = 2 \pi \rho v_r^3 r^2 \hspace{1cm} \eta_{K} = \frac{\rho v_r^3}{2F}.
\end{equation}
\end{subequations}
The efficiencies $\eta_q$ have been defined by normalizing each of the bulk wind fluxes $\dot{q}$ by the corresponding flux of the radiation field. We interpret $\eta$ as an efficiency because it is a measure of the coupling strength between the radiation field and the gas flow for that particular flux.

We plot each of the above fluxes and efficiencies in Fig. \ref{fig:efficiency_summary}. First we notice that for a given SED, kinematic fluxes scale with radiative flux $\dot{q} \sim F^{\alpha_q}$. Averaging over SEDs we find $\alpha_M, \alpha_n \sim 0.8$, $\alpha_p \sim 1$ and $\alpha_K \sim 1.4$. From the heating rate argument used to derive the critical flux we expect $T \sim F^2$. If there is an equipartition of kinetic and thermal energy we would therefore expect $v_r \sim F^{1}$. However, we see this dependence is in fact much weaker with $v_r \sim F^{0.2-0.4}$ Starting with the particle flux, each subsequent quantity carries an additional power of $v_r$ which from the basic scaling argument is expected to add an additional power of $F^{1/2}$. Comparing different SEDs we notice that there are two classes of SED - the main group and the BB SED which is somewhat of an outlier. We note that the BB SED lacks soft photons relative to all the other models, which leads to an overall decrease in the dynamical fluxes. This is a general trend with the other SEDs whereas the AGN SEDs have dynamical fluxes roughly an order of magnitude larger for fixed total flux. As a figure of merit $F_{\rm{soft}}/F \sim 99 \%$ for AGN, $70\%$ for Brem and $10^{-3} \%$ for BB. The BB fluxes resemble those of the other SEDs if we plot $F_{\rm{soft}}$ instead of total flux. This reflects the microscopic properties of the heating i.e. in the subsonic part of the wind, where dynamical fluxes are determined, it is the soft X-ray photons which play a dominant role in heating the wind. This can also be seen from the equilibrium curves where the base of the S-curve is determined by line cooling and heating due to photoionization, which are sensitive to photons from UV to soft X-rays.

The efficiencies also have a powerlaw scaling $\eta_{q} \sim F^{\beta_q}$ with $\beta_{M},\beta{n} \sim - 0.1$, $\beta_{p} \sim 0$ and $\beta_{K} \sim 0.4$. Unlike the dynamical fluxes, the efficiencies are relatively independent of the radiation flux. In particular, increasing the efficiency of mass, particle and momentum flux is easier to achieve by changing the external SED rather than increasing the radiative flux. The efficiency in driving the wind is more complex than the basic parameters characterizing the radiation field. For instance 5 models have the same X-ray temperature (mBl, Bl, Brem, BB and XRB1). Even ignoring (BB), the outlying SED,  which is much weaker in soft photons, the remaining SEDs still show a spread of roughly half an order of magnitude in efficiency. 

This suggests that accurately modeling the gross wind properties requires accurate modeling of the radiation field. In particular, simply knowing the X-ray temperature  and total flux is insufficient because as we have shown, efficiencies are weakly dependent on flux. For a fixed X-ray temperture, the efficiency can vary by almost an order of magnitude. The dynamical fluxes at large distances are a consequence of those same fluxes at the sonic point. However, because the sonic point occurs far away from the Compton radius, the primary heating mechanism is not Compton heating but rather photoionization. This heating rate is sensitive to the soft X-ray part of the radiation field and therefore a careful modeling of the gas is necessary.  

\subsection{Absorption Measure Distribution}
\label{sec:AMD}

%%%%%%%%%%%%%%%%%%%%%%%%%%%%%%%%%%% Efficiencies
\begin{figure}
                \centering
                \includegraphics[width=0.45\textwidth]{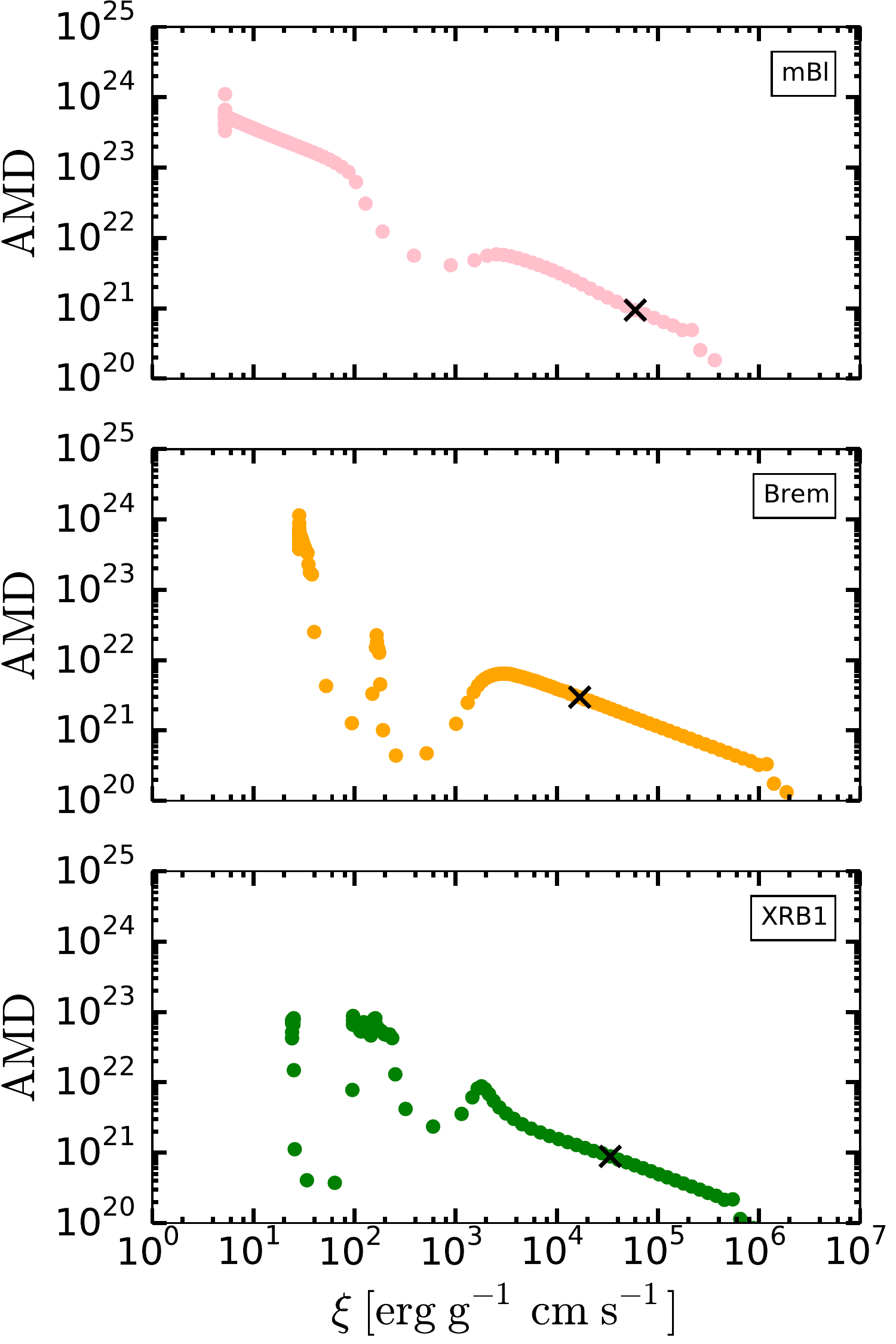}
        \caption{AMD for mBl (pink), Brem (orange) and XRB1 (green) models for flux $F/F_{\rm{cr}} = 10\%$ corresponding to the blue colored points in Fig. 2. The X indicates the sonic point. Dips occur when there is rapid heating of the flow. This adjustment to the radiation field happens in the subsonic regions. The above models have the same $T_X$ but exhibit qualitatively different behaviour. mBl has a dip despite the equilibrium curve being stable. Brem has two closely spaced dips from a TI and when adiabatic cooling becomes important. XRB1 has two dips for the same reason but has an intermediate temperature stable phase between these two dips.              
}
\label{fig:AMD_summary}
\end{figure} 
%%%%%%%%%%%%%%%%%%%%%%%%%%%%%%%%%%%

One may test the validity of theoretical wind solution by comparing wind densities, velocities and photoionization with observed values. However, in addition to finding that solutions have gas in the observed $(\rho, v_r, \xi)$ parameter space one should ensure that a sufficiently high column density of gas with those properties is present to in fact be observable. One such measure is the Absorption Measure Distribution (AMD; Holczer, Behar \& Kaspi 2007) which measures the absorbing column of gas at a particular ionization state along a line of sight. It is defined as
\begin{equation}
AMD = \frac{dN_H}{d(log \ \xi)} = - n \lambda_{n},
\end{equation}
where $N_H = \int n \ dr$ is the total hydrogen column density along the line of sight and $\lambda_n = (d \ln n/ d r)^{-1}$ is the characteristic length for particle number variations and we have used our uniform ionizing flux model (see eq. (\ref{eq:xi_def})). This is observationally interesting because particular lines will only be absorbed by a gas in a particular ionization state (see for example Adhikari et al. 2015). However, if the gas density or characteristic length or both are too low any such spectral features will be absent. We emphasize that important comparisons can be made to observations by detecting individual lines at a particular ionization state and that comparison with the entire distribution is unnecessary.

We find that all our models exhibit dips in the AMD where they experience rapid heating. This can occur in two ways - 1) when the equilibrium curve is formally unstable or 2) dynamical processes in the flow become important, triggering heating. The AMD behaves qualitatively differently in these different flows. In the case of a formally stable equilibrium curve (mBl), a dip occurs when dynamical processes become important and the flow falls out of equilibrium. At large radii the AMD follows a powerlaw, largely independent of any dips at smaller radii. The thermal instability leads to sharper dips because of the larger heating rates experienced by the gas. After the thermal instability, the gas can reach a new stable intermediate temperature state (see XRB1 for example).  The AMD then reaches a level commensurate to where it was before the unstable zone. If no such phase exists (see Brem for example) then the AMD will remain small at these ionizations.  This is because along the equilibrium curve the dynamical time scales are less important so higher densities of the gas can exist. The AMD can therefore probe several interesting features of the equilibrium curve. It can tell us about what parts of the gas are thermally unstable and also indicate that parts of the gas exist along an intermediate stable temperature. These predictions require both accurate modeling of the heating rates (to find the thermally unstable regions) and hydrodynamic flow (to see what parts of the equilibrium curve are probed).

\section{Discussion}
\label{sec:discussion}
Using the photionization code \textsc{XSTAR} we used both idealized and observed SEDs to generate the heating/cooling rates of a gas as a function of temperature and ionization parameter. We implemented this into a heating module in the hydrodynamics code 	\textsc{Athena++} to model spherically symmetric, thermally driven winds. The general behaviour and evolution can be characterized as follows.

\textbf{1)} We find stationary solutions that follow the classic picture of thermally driven winds - energy is added to the base of the wind in the subsonic region to overcome the gravitational potential and to accelerate the wind to supersonic velocities. The mass loss rate is determined by the amount of energy deposited at and below the sonic point, whereas other properties such as the wind velocity and energy flux are determined by the total amount of energy deposited in the supersonic part of the flow.

\textbf{2)} The mass, particle, momentum and kinetic energy fluxes are functions of the radiation field flux and the nature of the radiation field. The efficiency of driving the wind is a weak function of radiation flux but strongly dependent on the details of the SED.

\textbf{3)} The AMD provides a useful tool for determining the behaviour of the gas. When the gas experiences rapid heating there is a steep drop in the AMD. If gas exists at a stable intermediate temperature the AMD will have a bump at that ionization state.

To lowest order, one can assume that heating is dominated by Compton processes and the heating rate is simply parametrized by the X-ray temperature. This is a valid approximation at high temperatures and correctly predicts that the flow heats to the Compton temperature for a critical flux of incident radiation. However, at low temperatures, $T \ll T_X$, Compton processes are subdominant to photoionization heating and cooling due to line and free-free emission. Therefore the X-ray temperature is insufficient to predict the state of the gas. One may then go one step further and compute the equilibrium curve for a given SED. The state of the flow is approximated by traversing the equilibrium curve. Regions where the equilibrium temperature is very sensitive to the photoionization parameter indicate highly efficient net heating that can lead even to catastrophic heating with TI being an example.  A high energy deposition rate results in a flow acceleration and that in turn leads to adiabatic cooling. When the flow traverses another such region, as is possible for sufficiently high illuminating flux, it experiences another phase of acceleration. Consequently, the state of the flow is not properly captured by the radiative equilibrium solution. Our simulations show that falling out of radiative equilibrium has several interesting effects on the flow and these have important observational consequences.

Let us first compare how models with the lowest level of complexity behave. The mBl, Bl, Brem and XRB1 models all have the same $T_X = 2.8 \times 10^{7} ~ \rm{K}$. They also begin with the same gravitational and thermal energy at the base of the wind. However they have nearly an order of magnitude difference in the various fluxes and corresponding efficiencies as shown in Fig. \ref{fig:efficiency_summary}. For all cases, the sonic point occurs at a temperature $T \ll T_X$, hence the relevant temperature scale for determining mass flux at the critical point is not the X-ray temperature. Models with the same X-ray temperature therefore will not necesarily have similar mass flux, as our simulations show.  The other wind fluxes are proportional to the mass flux times additional powers of the velocity $v$. Therefore they differ by roughly an order of mangitude as well between the various models.

What about a model where we simply traverse the equilibrium curve? This is equivalent to assuming we provide a large enough flux for the radiative processes to dominate over any hydrodynamics. As a proxy for this model we will use the critical luminosity cases. However, as was shown in Section \ref{sec:bulk}  the various wind fluxes have power-law scaling $ \dot{q} \sim F^{0.8-1.4}$ and the efficiencies $\eta_{q} \sim F^{-0.1-0.4}$. Furthermore, it fails to predict when the flow falls out of equilibrium. This means that certain intermediate stable temperature branches which would naively be expected to exist from the equilibrium curve might never be visited because insufficient flux is heating the flow. This is observationally one of the most interesting features of this type of system because it provides a link between the macroscopic gas properties (temperature and ionization parameter) and the underlying microphysics.

In high flux cases, the flow has more opportunities to traverse steep portions of the S curve as it explores a larger portion of it. Though not necessarily formally unstable to the TI, these regions where temperature varies strongly with ionization parameter lead to an acceleration of the flow. High flux cases can therefore experience multiple phases of acceleration, provided they remain in radiative equilibrium and the S curve features multiple steep regions. We found this was the case with the mBl SED where in the highest flux runs had two stages of acceleration. The AMD features two dips, indicating the ionization state of the gas when it accelerates.

We have assumed an optically thin wind, though a posteriori we can estimate the validity of this approximation. Except when $F = F_{\rm{cr}}$, all cases have column densities $N_H < 10^{24} \rm{g \ cm^{-2}}$ and therefore are optically thin throughout. The critical case is optically thin in the supersonic part. This indicates that the optically thin heating assumption is not always valid, however it can be used as a starting point for this type of calculation.

Our results suggest that complete spectral information is important when modeling radiative heating and thermally driven flows. This is especially imprtant for systems like AGN. Besides the Type 1 and Type 2 AGN cases, we may also interpret the BB case as an AGN which has nearby cold gas that absorbs the soft photons. The AGN1 and AGN2 cases have a difference of a few in their various wind fluxes and efficiencies. However the BB case is several orders of magnitude smaller, primarily due to highly suppressed soft photons. When modeling observed AGN outflows this is important because any uncertainty in the observed flux will result in an uncertainty in the modeled outflow.     

We found numerical artifacts affecting our solution in several ways. This suggests the need for new physics to properly resolve the flow. When the wind becomes non-adiabatic and falls of the equilibrium curve, there is a drop in the length scale associated with the dynamical variables. This becomes more pronounced at higher radiation fluxes and can lead to not finding a stationary numerical solution. Likewise when solutions are thermally unstable there is a decrease in the length scale of dynamical variables - if it falls below the grid scale the TI is not numerically well resolved. This cannot be adressed with additional resolution as the regions are geometrically thin. Future studies of wind solutions should include the effects of thermal conduction because of relatively steep temperature gradients (see for example Rozanska \& Czerny 1996).  

Our future studies will include 2D and 3D simulations of disc winds driven by the radiation force. In addition to a considerably more complex geometry and dynamics due to a rotating disc for instance, accurate modeling of the radiative heating is also critical. If the disc is irradited by a central source or the inner parts of the disc, the outer disc will heat. This affects the strength of the radiation force and therefore the radiatively driven wind. However it has been shown that shadowing of the source by an inner disc wind is an important effect in this type of setup (e.g., Proga, Stone \& Kallman 2000 and Higginbottom et al. 2014). If line driving is important, calculating the force multiplier resulting from the lines responsible for the heating/cooling as well as additional lines that are negligible energetically but important for momentum transfer. By using observed SEDs as an input to \textsc{XSTAR} we will model self-consistently the heating/cooling in the system. We also note that this method is quite general and can be applied to other radiatively heated systems for which an SED has been observed. For example one could consider the net heating of planetary atmospheres by the host star. 

\section*{Acknowledgements}
This work was supported by NASA under ATP grant NNX14AK44G. We thank Maria Diaz Trigo for providing the XRB SEDs and Missagh Mehdipour for providing the AGN SEDs used in our work. We also thank Tim Kallman for providing us with \textsc{XSTAR} and many valuable discussions.

\appendix
\section{Heating Function}
\label{sec:heating_appendix}

%%%%%%%%%%%%%%%%%%%%%%%%%%%%%%%%%%%
\begin{figure*}
                \centering
                \includegraphics[width=\textwidth]{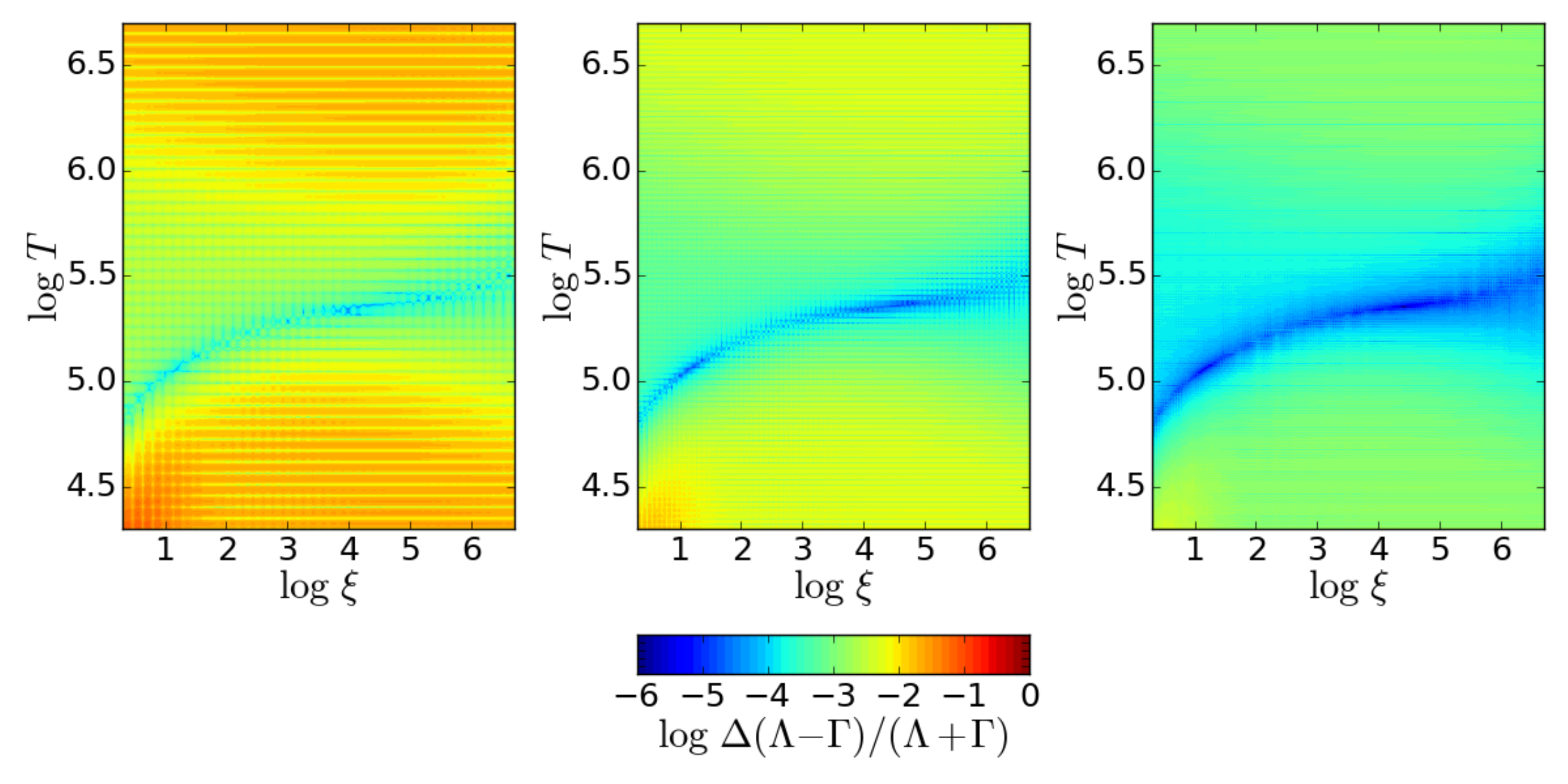}
        \caption{Relative percent error in bilinearly interpolated heating/cooling rate (\ref{eq:relative_heating_cooling}) as a funcion of position in $(T,\xi)$ phase space for tables of size $N \times N$ for $N$ = 50 (\textit{left}) , 100 (\textit{center}) \& 200 (\textit{right}).       
}
\label{fig:summary_xiT_modelA}
\end{figure*} 
%%%%%%%%%%%%%%%%%%%%%%%%%%%%%%%%%%%

%%%%%%%%%%%%%%%%%%%%%%%%%%%%%%%%%%%
\begin{figure}
                \centering
                \includegraphics[width=0.45\textwidth]{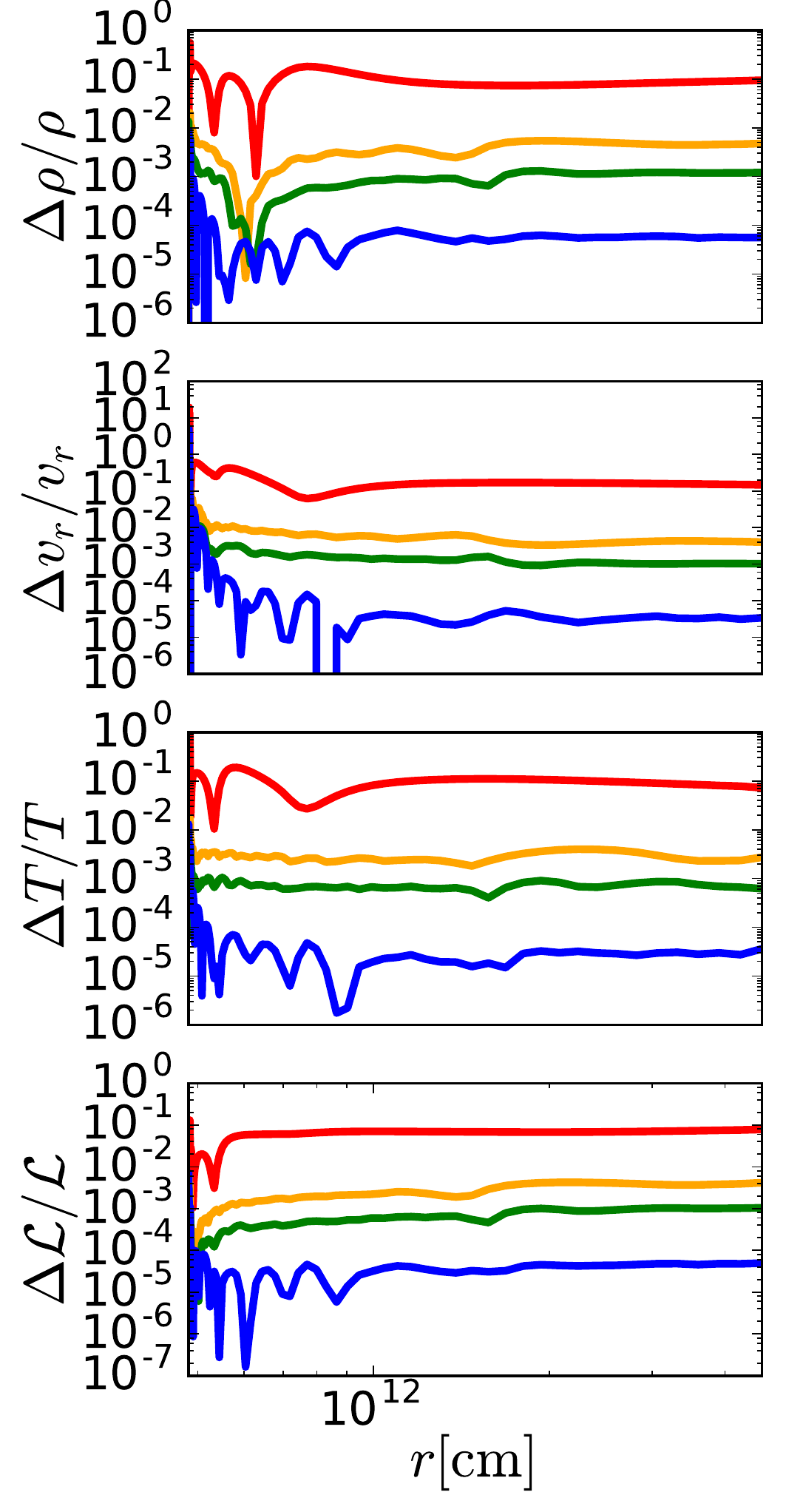}
        \caption{Relative error in stationary wind solutions in simulations using bilinearly interpolated compared to analytically computed heating/cooling rates for tables of size $N \times N$ for n = 10 (red), 50 (orange), 100 (green) \& 500 (blue).      
}
\label{fig:summary_stationary}
\end{figure} 
%%%%%%%%%%%%%%%%%%%%%%%%%%%%%%%%%%% 

Our 	\textsc{Athena++} simulation calculates radiative heating rates self consistently by interpolating the net heating rates precalculated using the XSTAR photoionization code for a range of temperatures and ionization parameters. We describe this interpolation scheme in Section \ref{sec:hcmodule_methods} and the results of various performance tests in Section \ref{sec:hcmodule_tests}.  

\subsection{Algorithm}
\label{sec:hcmodule_methods}
We calculate the net heating rate using XSTAR for an $M \times N$ logarithmically spaced grid $(\xi_i, T_j)$ in a domain $T_{\rm{min}} < T < T_{\rm{max}}$ and $\xi_{\rm{min}} < \xi < \xi_{\rm{max}}$. We take $T_{\rm{min}} = 5 \times 10^3  \rm{K}$, $T_{\rm{max}} = 10^7  \rm{K}$, $\xi_{\rm{min}} = 10^0 \ \rm{erg \ cm^{-3}}$ and $\xi_{\rm{max}} = 10^8 \ \rm{erg \ cm^{-3}}$. We will refer to this as our heating table. 

Within 	\textsc{Athena++}, at every cell and timestep, the net heating must be calculated. This requires two interpolations: \textbf{1)} In $\xi-T$ phase space to determine the nearest points in the heating table \textbf{2)} In T space to implicitly solve for the heating rate consistent with the hydro time step.  

\subsubsection{$\xi-T$ Space - Bilinear Interpolation}
When we require the heating rate for ionization and temperature $(\xi,T)$ we \textbf{1)} Determine the (i,j) location in the heating table for the four pairs of points $(\xi_i,T_j)$, $(\xi_{i+1},T_j)$, $(\xi_i,T_{j+1})$ and $(\xi_{i+1},T_{j+1})$ nearest to $(\xi,T)$. \textbf{2)} Bilinearly interpolate between these points to estimate $\mathcal{L}(\xi,T)$.

We have generated the heating table uniformally in logarithmic space according to 
\begin{equation}
\xi_i = i \left(\frac{\log_{10}\xi_{\rm{max}} - \log_{10}\xi_{\rm{min}}}{N}\right) + \log_{10}\xi_{\rm{min}},
\end{equation}
\begin{equation}
T_j = j \left(\frac{\log_{10}T_{\rm{max}} - \log_{10}T_{\rm{min}}}{M}\right) + \log_{10}T_{\rm{min}}.
\end{equation}     
Therefore when we are finding the indices $(i,j)$ we do not have to traverse the entire table we can simply invert the above expressions. We then use the standard bilinear interpolation scheme on these four points.

\subsubsection{$T$ Space - Backward Euler Interpolation}
We discretize the energy evolution equation (\ref{eq:energy}) under the assumption that changes in internal energy are due to the heating/cooling and express this in terms of the implicit relation
\begin{equation}
T^{n+1} = T^{n} - \Delta t \frac{(\gamma - 1) \mu m_p}{k} \mathcal{L}(T^{n+1},\xi).
\label{eq:backward_euler} 
\end{equation} 
This is the backward Euler method where the heating rate during the $n$th time step is calculated using the temperature at the $(n+1)$th time step. The alternative forward Euler method calculates the heating rate during the $n$th time step using the temperature at the $n$th time step.  The advantage of backward Euler method is that as the equilibrium temperature is reached the contribution from the heating will also be decreasing so equilibrium is reached more smoothly. The disadvantage is $T^{n+1}$ is now implicitly defined by (\ref{eq:backward_euler}) and a root finder is needed to solve for it. 

We implemented a two step bisection root finder. The process of finding a zero of a function $f(x)$ i.e solving for $x_0$ such that $f(x_0) = 0$ consists of two steps - \textbf{1)} bracketing the zero, i.e. finding numbers a,b such that $f(a)f(b) < 0$. A zero will then lie in the interval [a,b]. \textbf{2)} Refining this interval such that $b-a < \delta$ so that we can say that we have located the location of the zero $x_0$ to a precision $\delta$. 
 
Step \textbf{1)} consists of checking for the existence of a root in progresively larger intervals $[T_a, T_b]$ where $T_a = \rm{max}\left\{T/\alpha^n,T_{\rm{min}} \right\}$ and $T_b = \rm{min}\left\{T \alpha^n,T_{\rm{max}} \right\}$  where $\alpha = 1.05$ and T is the current temperture. We then check the condition $f(T_a)f(T_b) < 0$ and if it is not satisfied continue decreasing (increasing) $T_a$ ($T_b$) until it is satisfied. Step \textbf{2)} uses the \textbf{zbrent} rootfinding algorithm (Press et al. 1992) to solve for the root to an accuracy $\delta = 100 \rm{K}$.

We compared how both methods performed when calculating the heating (cooling) of a box of uniformally ionized gas at temperature below (above) its equilibrium temperature. The backwards Euler method did not oscillate above/below the equilibrium temperature as with forward Euler. This is an important property because we expect wind solutions to be at equilibrium near their base. A comparable level of performance by the forward Euler method requires decreasing the timestep, which in practice entails more computational expense than employing the root finder.

\subsection{Tests}
\label{sec:hcmodule_tests}

To test our algorithm and implementation we use the analytic heating/cooling rates given by (\ref{eq:heating_cooling}) (Blondin 1994). We compare simulation results where this forumula is used explicitly and simulations where our algorithm is used to interpolate over heating/cooling rates calculated on an $N \times N$ grid. 

We perform two main tests \textbf{1)} Compute the relative error in the heating/cooling rate at different locations in the ($T,\xi$) phase space to test the bilinear interpolation scheme. \textbf{2)} Compute the relative error in the kinematic quantities and heating rates for radiatively heated 1D stationary winds in 	\textsc{Athena++}.

\subsubsection{Bilinear Interpolation}

We consider $N \times N$ grids of points $(\xi_i,T_j)$, logarithmically spaced in the $(\xi,T)$ plane. We consider grids of size N = 50, 100 \& 200. We then evaluate the heating/cooling rate in a logarithmically spaced grid of size $1000 \times 1000$ in the $(\xi,T)$ plane using the analytic expression (\ref{eq:heating_cooling}) and by using our bilinear interpolation scheme on the $N \times N$ grid. We ensure this grid is shifted relative to the grid used to calculate heating rates so we are testing the interpolation scheme. 

We show results for N = 50, 100 \& 200 in Fig. \ref{fig:summary_xiT_modelA}. We have plotted the relative error in the numerically interpolated net heating rate (subscript n), normalized to the average analytic heating and cooling (subscript a)
\begin{equation}
\frac{\Delta (\Lambda - \Gamma)}{\Lambda + \Gamma} = 2 \frac{|(\Lambda - \Gamma)_{a} - (\Lambda - \Gamma)_{n}|}{|\Lambda + \Gamma|_{a}}.
\label{eq:relative_heating_cooling}
\end{equation}
We normalize the error in this way to not exagerate the importance of numerical errors near or at equilibrium where $(\Lambda - \Gamma) \rightarrow 0$.  

We note that the error contours are stratified, particularly at small N. This is because near the calculated points $(\xi_i, T_j)$ the bilinear approximation will perform best. As we distance ourselves from this point and before we approach the next point on the grid the approximation will worsen.

We see that if we want to achieve roughly 1\% accuracy we need roughly a $50 \times 50$ grid. In order to achieve 0.1\% accuracy we need a grid of $100 \times 100$.

\subsubsection{Stationary Solution}

We consider $N \times N$ grids of points $(\xi_i,T_j)$, logarithmically spaced in the $(\xi,T)$ plane. We consider grids of size N = 10, 50, 100 \& 500. We perform 	\textsc{Athena++} simulations where the heating/cooling rate is determined by the bilinear interpolation scheme on this grid. After 	\textsc{Athena++} has found a stationary solution, where the hydrodynamic time step $dt$ becomes constant, we compare these to the stationary solutions found when heating/cooling is computed analytically. We plot the relative \% error in the late time density, velocity, temperature and heating rate in Fig. (\ref{fig:summary_stationary}).

Relative errors are large near the star where the solution is subsonic. However, past the sonic point the relative error shows a clear trend in the accuracy of the simulations. In particular a $50 \times 50$ grid yields roughly 1\% accuracy while a $100 \times 100$ grid yields a 0.1\% accuracy. Interestingly the accuracy in the final solution is comparable to the accuracy in the $(\xi,T)$ plane discussed above. 

We note that the solution follows the equilibrium curve near the star before falling off and staying slightly below it in the heating regime. The location where the wind is no longer in equilibrium corresponds to the radius at which the relative \% error converges to its large radius value. It makes sense that inside this radius the \% error has large variations, because the analytic value should be near zero and therefore the relative errors are amplified.

\end{document}